\newcommand{\Tr}{\mathop{\mathrm{Tr}}\nolimits}
\documentclass[osajnl,twocolumn]{revtex4}

\usepackage{amsfonts,amssymb,amsbsy,bm,graphicx}

\begin{document}

\title{Integral estimator of broadband omnidirectionality}

\author{Alberto G. Barriuso, Juan J. Monz\'on, and Luis L. S\'anchez-Soto}
\affiliation{Departamento de \'Optica, Facultad de F\'{\i}sica,
Universidad Complutense, 28040 Madrid, Spain}

\author{\'Angel Felipe}
\affiliation{Departamento de Estad\'{\i}stica
e Investigaci\'on Operativa I, Facultad de Matem\'aticas,
Universidad Complutense, 28040 Madrid, Spain}

\begin{abstract}
By using the notion of wavelength- and angle-averaged
reflectance, we assess in a systematic way the performance
of finite omnidirectional reflectors. We put forward how
this concept can be employed to optimize omnidirectional
capabilities. We also apply it to give an alternate
meaningful characterization of the bandwidth of these
systems.
\end{abstract}


\maketitle

Floquet-Bloch theory warrants that the behavior of
periodically stratified media is determined by the
trace of the transfer matrix of the basic period.
Indeed, whenever the magnitude of this trace is
greater than 2, no waves  propagate in the structure
and then a stop band appears.

In the context of electromagnetic optics, this
notion is at the basis of photonic crystals~\cite{Dowling}
(that is, one-dimensional periodic layered
structures), which have been attracting a
lot of attention because of their amazing property
of acting as omnidirectional reflectors (ODRs):
they reflect light at any polarization, any
incidence angle, and over a wide range of
wavelengths~\cite{Yablonovitch1987,John1987,Fink1998,Dowling1998,Chigrin1999}.

Although there are a number of approaches for
ensuring a trace greater than 2 in the basic
period~\cite{Liu1997,Macia1998,Cojocaru2001,Lusk2001,Peng2002},
the most feasible design involves two materials
with refractive indices as different as possible.
Such a bilayer system is usually designed at
quarter-wave thickness (at normal incidence),
which is enough to guarantee
ODR~\cite{Southwell1999,Chigrin1999a,Lekner2000}.
However, this assumes perfect periodicity and so
requires the system to be strictly infinite. Of
course, this is unattainable in practice and one
is led to consider stacks of $N$ periods, which
are often appropriately called finite periodic
structures~\cite{Lekner1994}. One can rightly
argue that when $N$ is high enough (say, 50 or
more), there should be no noticeable differences
with the ideal infinite case~\cite{Lusk2005}. But
there are commercial ODR designs considering only
very few  periods~\cite{ssp} and, in such a
situation, the optimization of the basic
period deserves a careful and in-depth study.

To shed light on this issue it is essential
to quantify the ODR performance in a manner
that permits unambiguous comparison between
different structures. We hold to previous
suggestions~\cite{Yonte2004,Barriuso2005},
but to take into due consideration the key role
of the bandwidth, we propose here to average the
reflectance over all the incidence angles and
all the wavelengths in the spectral range.
With this tool at hand, we revisit finite ODRs
and characterize their properties, addressing,
as a side product, a proper picture of
the omnidirectional bandwidth for these
systems.

We begin by briefly recalling some background
concepts. The basic period of the finite ODR
consists of a double layer made of materials
with refractive indices $(n_L, n_H)$ and
thicknesses $(d_L, d_H)$, respectively. The
material $L$ has a low refractive index, while
$H$ is of a high refractive index. To characterize
the optical response we employ the transfer matrix,
which can be computed as $\mathsf{M}  = \mathsf{M}_L \,
\mathsf{M}_H$, where $\mathsf{M}_L$ and
$\mathsf{M}_H$ are the transfer matrices
of each layer, whose standard form can be
found in any textbook~\cite{Yeh1988}.

As we have mentioned before, band gaps appear
whenever the trace of the basic period satisfies
\begin{equation}
\label{TrODR}
| \Tr ( \mathsf{M} ) | \ge 2 \, .
\end{equation}
This condition should be worked out for both basic
polarizations. However, it is known that whenever
Eq.~(\ref{TrODR}) is fulfilled for $p$ polarization,
it is always true also for $s$ polarization. The
$p$-polarization bands are more stringent than the
corresponding $s$-polarization ones~\cite{Lekner1987}
and we thus restrict our study to the former.

We next consider an $N$-period structure whose
basic cell is precisely the bilayer $LH$. We
denote this as $[LH]^N$ and its overall transfer
matrix is simply $\mathsf{M}^N,$ from which
calculating its reflectance $\mathcal{R}^{(N)}$
is straightforward. According to our previous
discussion,  we average over all the incidence
angles and all the wavelengths in the spectral
interval $\Delta \lambda = \lambda_{\mathrm{max}}
- \lambda_{\mathrm{min}}$ of interest
\begin{equation}
\label{volumen}
\overline{\mathcal{R}}^{(N)} =
\frac{1}{\Delta \lambda}
\int_{\lambda_{\mathrm{min}}}^{\lambda_{\mathrm{max}}}
\left (
\frac{2}{\pi}
\int_{0}^{\pi/2} \mathcal{R}^{(N)} \ d \theta \right )
d \lambda \, ,
\end{equation}
and take this as an appropriate figure of merit to
assess the performance as an ODR. Once the materials
have been chosen, $\overline{\mathcal{R}}^{(N)}$ is
a function exclusively of the layer thicknesses.

As a case study, we take the materials to be cryolite
(Na$_3$AlF$_6$) and zinc selenide (ZnSe), with refractive
indices $n_L = 1.34$ and $n_H = 2.568$, respectively, at
the wavelength $\lambda = 0.65 \ \mu$m. The spectral window
considered is from $\lambda_{\mathrm{min}} = 0.5 \ \mu$m
to $\lambda_{\mathrm{max}} = 0.8 \ \mu$m. In this range,
the refractive index of the cryolite can be considered,
to a good approximation, as constant, while for the zinc
selenide we use the Sellmeier dispersion equation $n_H^2
(\lambda ) = 4 + 1.9 \lambda^2/ [\lambda^2- (0.336)^2]$,
where $\lambda$ is expressed in microns.

Many commercial packages are available to  perform layer
optimization. In the case of ODR, common methods
optimize a merit function that (quadratically)
measures how the calculated reflectance separates
from unity (ideal target) at some definite angles
and at some definite wavelengths.  For example, TFCALC
uses needle optimization to find the best thicknesses
for such a merit function. We have preferred, however,
to implement a gradient-based modified quasi-Newton
algorithm (using the Fortran NAG libraries), for
its consistency with the problem investigated, which
is continuous.

\begin{figure}
\centering
\resizebox{0.90\columnwidth}{!}{\includegraphics{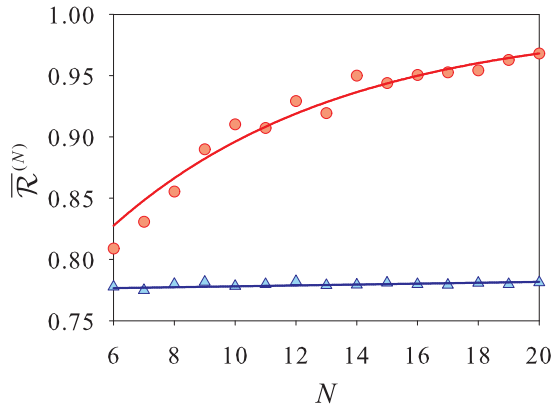}}
\caption{(Color on line) Averaged reflectance
$\overline{\mathcal{R}}^{(N)}$ in terms of number
of periods for the optimized $[LH]^N$ systems (blue triangles)
and for the same system when all the thicknesses can
be different (red circles). Solid lines represent the
fittings to these points.}
\end{figure}

We have optimized the system $[LH]^N$ for $N$ running
from 5 to 20. When the thicknesses are kept equal by
pairs, so as the structure retains its periodicity,
we find thicknesses distributed around $d_L \simeq
145$~nm and $d_H \simeq 60$~nm. The variation of
$\overline{\mathcal{R}}^{(N)}$ with $N$ is shown
in Fig.~1, and can be numerically fitted to
$\overline{\mathcal{R}}^{(N)} = 0.774 + 0.365
\times 10^{-3} \ N$; i.e., a linear dependence
with an extremely small slope. We have also worked
out the instance when all the thicknesses may vary
independently (though now the system is not strictly
periodic). The optimum thicknesses oscillate a lot,
without a definite pattern (we do not list all of
these values because of space limitation). However,
as is clear from Fig.~1,  $\overline{\mathcal{R}}^{(N)}$
shows an exponential increasing that can be suitably
represented by $\overline{\mathcal{R}}^{(N)} = 0.625 +
0.369 [1 - \exp(- 0.132 N )]$.

\begin{figure}
\begin{center}
\centering
\resizebox{0.90\columnwidth}{!}{\includegraphics{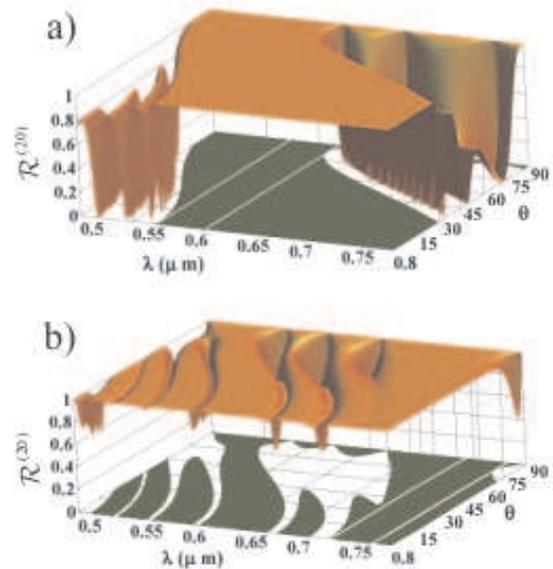}}
\caption{(Color on line) Plot of the reflectance
$\mathcal{R}^{(20)}$ as a function of the incidence
angle $\theta$ (in degrees) and the wavelength $\lambda$
(in $\mu$m). In a) we have the $[LH]^{20}$ system, while
in b) the thicknesses are allowed to be different. At
the bottom plane, we show the contour plots corresponding
to a reflectance 0.99. The white lines delimit the zones
in which this reflectance is greater than 0.99 for all
the angles of incidence.}
\end{center}
\end{figure}

To gain further insights into these striking
differences, in Fig.~2 we have plotted the
reflectance $\mathcal{R}^{(20)}$ for the optimum
thicknesses~\cite{rep} corresponding to $N = 20$
as a function of the angle of incidence $\theta$
and the wavelength $\lambda$. At the bottom plane
we have also included  the contours of the regions
where this reflectance is greater than 0.99. While
for the $[LH]^{20}$ system, the top zone looks
quite flat and very close to unity, the dips are
very deep indeed. On the contrary, when we allow
for different thicknesses, the top zone presents
small ripples, but the dips are much less pronounced.
Curiously enough, the $[LH]^{20}$ system gives a
wider region lying above the 0.99 level. We have
also marked two stripes (within parallel lines)
in which the reflectance is greater than
0.99 for all the angles of incidence (and that,
roughly speaking, could be identified with stop
bands). Although of similar extension, they lie
in quite different spectral ranges. We have
repeated these calculations with other values
of $N$, observing essentially the same kind of
behavior.

\begin{figure}
\centering
\resizebox{0.95\columnwidth}{!}{\includegraphics{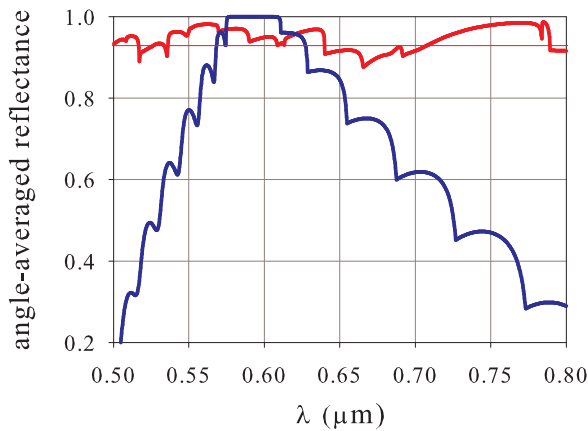}}
\caption{(Color on line) Angle-averaged reflectance
in terms of the wavelength $\lambda$ for the
optimized $[LH]^{20}$ system (blue) and for the
analogous system with different thicknesses (red).
The line of reflectance equal to 0.95 is also shown.}
\end{figure}

To proceed further, in Fig.~3 we have represented the
angle-averaged reflectance [i.e., the magnitude in
parentheses in Eq.~(\ref{volumen})] for the same two
systems as in Fig.~2, in terms of the wavelength $\lambda$.
Obviously, the area under the curve is precisely
$\overline{\mathcal{R}}^{(20)}$. It is clear that for
different thicknesses, this area is considerably bigger
(so, it is really an optimum), while the $[LH]^{20}$
is better behaved in a narrow range going from  0.57
$\mu$m to around 0.62 $\mu$m (which matches well
with the stop band shown in Fig.~2.a). In other
words, the optimum system remarkably improves the
behavior in the spectral wings, while it little
affects  (or even deteriorates) the behavior in
the ``good" central region. The ripples in both
curves are caused by the dips appearing in Fig.~2
for each line of $\lambda$ constant.

In this respect, we wish to note that, in our opinion,
the notion of bandwidth becomes fuzzy for ODRs.
Usually~\cite{Yeh1988}, it is defined as $ \delta
\lambda = \lambda_{+} - \lambda_{-}$ [and
sometimes~\cite{Southwell1999} normalized to the
central wavelength $( \lambda_{+} + \lambda_{-})/2$],
where $\lambda_{+}$ and $\lambda_{-}$ are the longer-
and shorter-wavelength edges for given ODR bands
[i.e., the two solutions of Eq.~(\ref{TrODR})]. This is
meaningful in the limit $N \rightarrow \infty$,  when
these band edges make unambiguous sense, but fails for
our more realistic situation. Other authors~\cite{Orfanidis2004}
note that the common reflecting band for both polarizations
and for angles up to a given $\theta_{\mathrm{M}}$
(which is defined by convention)  is
$[\lambda_{-} (\theta_{\mathrm{M}}), \lambda_{+} (0)]$
and the corresponding bandwidth $\delta \lambda =
\lambda_{+} (0) - \lambda_{-} (\theta_{\mathrm{M}})$.
Again, this kind of definition assumes the existence
of a full band, which is only true for the strictly
periodic case.

We emphasize that all the relevant information about
omnidirectionality is contained in the angle-averaged
reflectance. Furthermore, being an integral estimator,
it does not rely on the values of the reflectance
at some specific relevant angles. For this reason,
a sensible choice for defining the bandwidth is precisely
the spectral range(s) for which this angle-averaged
reflectance is bigger than a fixed threshold value.
For example, if we set this value to, say 0.95, it
is evident in Fig.~3 that the bandwidth for the
$[LH]^{20}$ system is always poor.

To sum up in a few words, we have exploited the
notion of wavelength- and angle-averaged reflectance
to explore in a systematic way the performance of finite
ODRs. Needless to say, our approach is general and can be
applied to other materials and other spectral regions.

This work has been supported by the Spanish Research
Agency Grant FIS2005-06714.


\begin{thebibliography}{10}
\bibitem{rep}
To ensure the reproducibility of our results,
we quote here the optimum thicknesses (expressed
in nm). For the $[LH]^{20}$ they are $d_L = 144$ and
$d_H = 59$. When all of them are different, we have
for the $L$-medium: 190, 111, 184, 99, 164, 145, 173,
180, 118, 132, 143, 134, 193, 101, 155, 122, 100, 165, 91,
110. For the $H$ medium they are 58, 141, 92, 140,
54, 122, 88, 71, 79, 46, 47, 72, 125, 143, 87, 112, 106,
59, 120, 116.
\end{thebibliography}

\end{document}